\documentclass[twocolumn,amsmath,amssymb,prl]{revtex4-1}
\usepackage{graphicx}
\usepackage{dcolumn}
\usepackage{bm}

\begin{document}

\title{Ultrafast strain engineering in complex oxide heterostructures}
\author{A.D. Caviglia$^{1}$, R. Scherwitzl$^{3}$, P. Popovich$^{1}$, W. Hu$^{1}$, H. Bromberger$^{1}$, R. Singla$^{1}$, M. Mitrano$^{1}$, M. C. Hoffmann$^{1}$, S. Kaiser$^{1}$, P. Zubko$^{3}$, S. Gariglio$^{3}$, J.-M. Triscone$^{3}$, M. F\"{o}rst$^{1}$, A. Cavalleri$^{1,2}$}
\affiliation{ $^{1}$Max-Planck Research Group for Structural Dynamics - Center for Free Electron Laser Science, University of Hamburg, Germany}
\affiliation{ $^{2}$Department of Physics, Clarendon Laboratory, University of Oxford, UK}
\affiliation{ $^{3}$D\'epartement de Physique de la Mati\`ere Condens\'ee, University of Geneva, 24 Quai Ernest-Ansermet, 1211 Gen\`eve 4, Switzerland}
\email{Andrea.Caviglia@mpsd.cfel.de}
\date{\today}

\begin{abstract}
We report on ultrafast optical experiments in which femtosecond mid-infrared radiation is used to excite the lattice of complex oxide heterostructures. By tuning the excitation energy to a vibrational mode of the substrate, a long-lived five-order-of-magnitude increase of the electrical conductivity of NdNiO$_3$ epitaxial thin films is observed as a structural distortion propagates across the interface. Vibrational excitation, extended here to a wide class of heterostructures and interfaces, may be conducive to new strategies for electronic phase control at THz repetition rates.
\end{abstract}

\maketitle
Complex oxide heterostructures have emerged as multifunctional materials of striking flexibility, in which unconventional electronic phases can be realised by engineering the strain field across interfaces \cite{Haeni:2004uq,doi:10.1146/annurev.matsci.37.061206.113016,Zeches13112009,Lee:2010fk,doi:10.1146/annurev-conmatphys-062910-140445}. This same mechanical coupling is also expected to be effective on the ultrafast timescale, and could be exploited for the dynamic control of materials properties. Here, we demonstrate that a large-amplitude mid-infrared field, made resonant with a stretching mode of the substrate, can switch the electronic properties of a thin film across an interface. Exploiting dynamic strain propagation between different components of a heterostructure, insulating antiferromagnetic NdNiO$_3$ is driven through a prompt, five-order-of-magnitude increase of the electrical conductivity, with resonant frequency and susceptibility that is controlled by choice of the substrate material.

Many of the functional properties of $AB$O$_3$ perovskite oxides are extremely sensitive to rotation and tilting of the $B$O$_6$ octahedra, which control the hopping amplitudes and the exchange interaction through the $B$-O-$B$ bond angle and length. It follows that one can engineer the electronic and magnetic properties by designing or actively controlling such distortions  \cite{PhysRevLett.105.227203,PhysRevB.83.153411,PhysRevLett.107.116805}, or by coupling them to other structural instabilities \cite{Bousquet:2008fk,PhysRevLett.106.107204}. One class of materials in which octahedral rotations and distortions have been linked to changes in the electronic structure are the rare earth nickelates, which display a sharp transition from a high-temperature metallic to a low-temperature insulating state. In the bulk, this electronic phase transition is accompanied by a structural change from an orthorhombic (Pbnm) to a monoclinic (P2$_1$/n) symmetry with an increase in the Ni-O-Ni bond bending and the appearance of a charge density wave. Additionally, charge disproportionation between adjacent Ni sites is associated with different Ni-O bond lengths \cite{0953-8984-9-8-003,doi:10.1080/01411590801992463}. At low temperatures, the nickelates also possess an unusual antiferromagnetic spin arrangement \cite{PhysRevB.73.100409}.

The metal-insulator transition in nickelates has been investigated so far using static experimental techniques that directly affect the electronic bandwidth through the Ni-O-Ni bond angle, such as chemical \cite{PhysRevB.45.8209,PhysRevB.61.1756} and hydrostatic pressure \cite{PhysRevB.47.12357,PhysRevB.52.9248} or epitaxial strain \cite{PhysRevB.62.7892}. Recently it was also shown that electrostatic fields can be used to control the metal-insulator transition in these materials \cite{ADMA:ADMA201003241}. Ultrafast optical experiments have measured carrier relaxation \cite{PhysRevB.76.165107} at low irradiation levels, insufficient to drive a photo-induced phase transition.

In this work we use intense coherent femtosecond mid-infrared pulses to control the lattice structure and with it the electronic properties along a non equilibrium path. Indeed strong vibrational excitation \cite{Forst:2011fk} is capable of inducing electronic phase transitions, as demonstrated in a recent series of experiments on cuprates \cite{Fausti14012011} and manganites \cite{Rini:2007mz,PhysRevLett.101.197404} where transient superconductivity and metallicity were triggered. In all of these cases the $B$-O stretching mode ($\sim$70-80\,meV) of the crystal structure was excited. Here we use epitaxy to mechanically couple NdNiO$_{3}$ to a substrate, whose $B$-O stretching mode is selectively driven. 

\begin{figure}
\begin{center}
\includegraphics[scale=0.20]{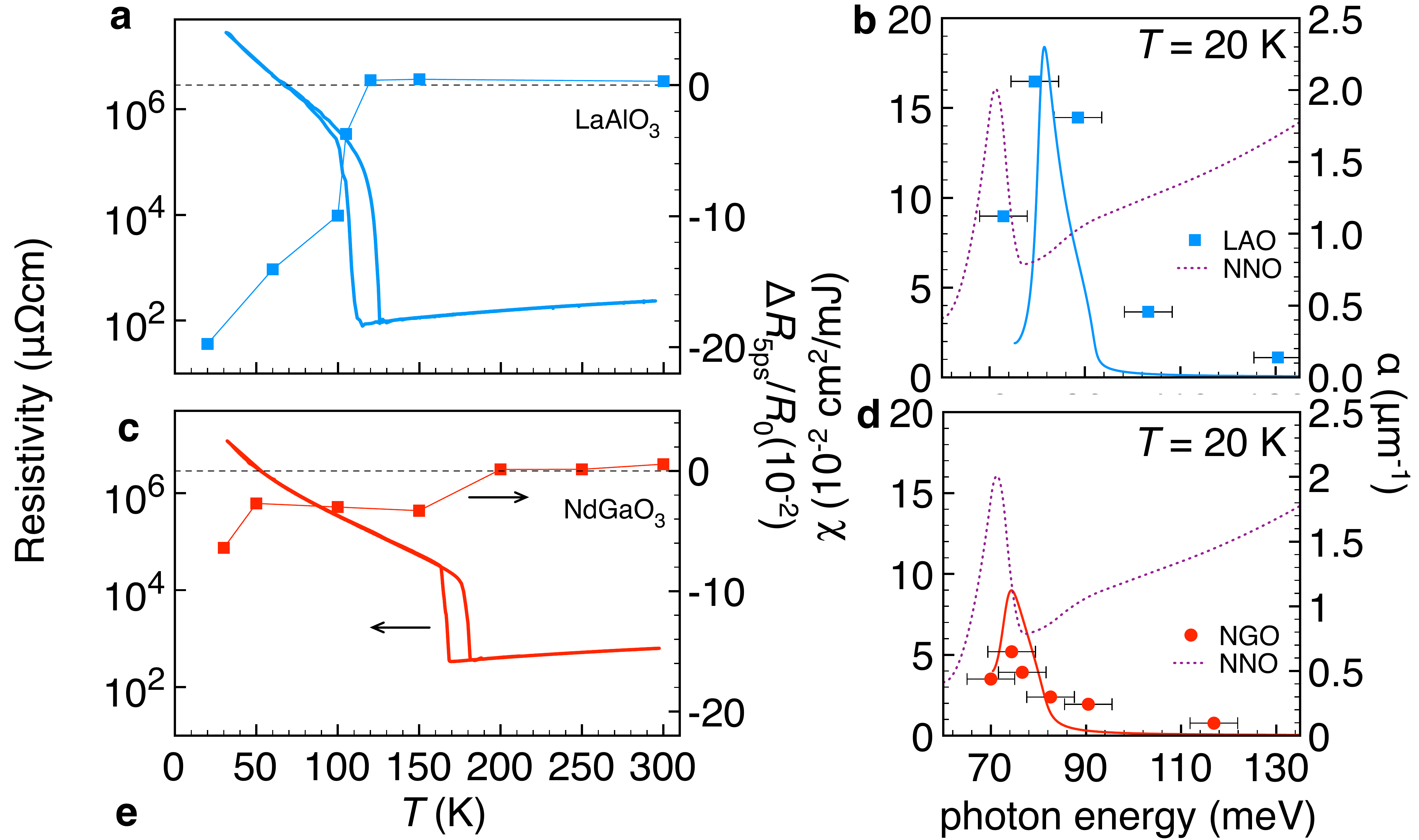}
\includegraphics[scale=0.06]{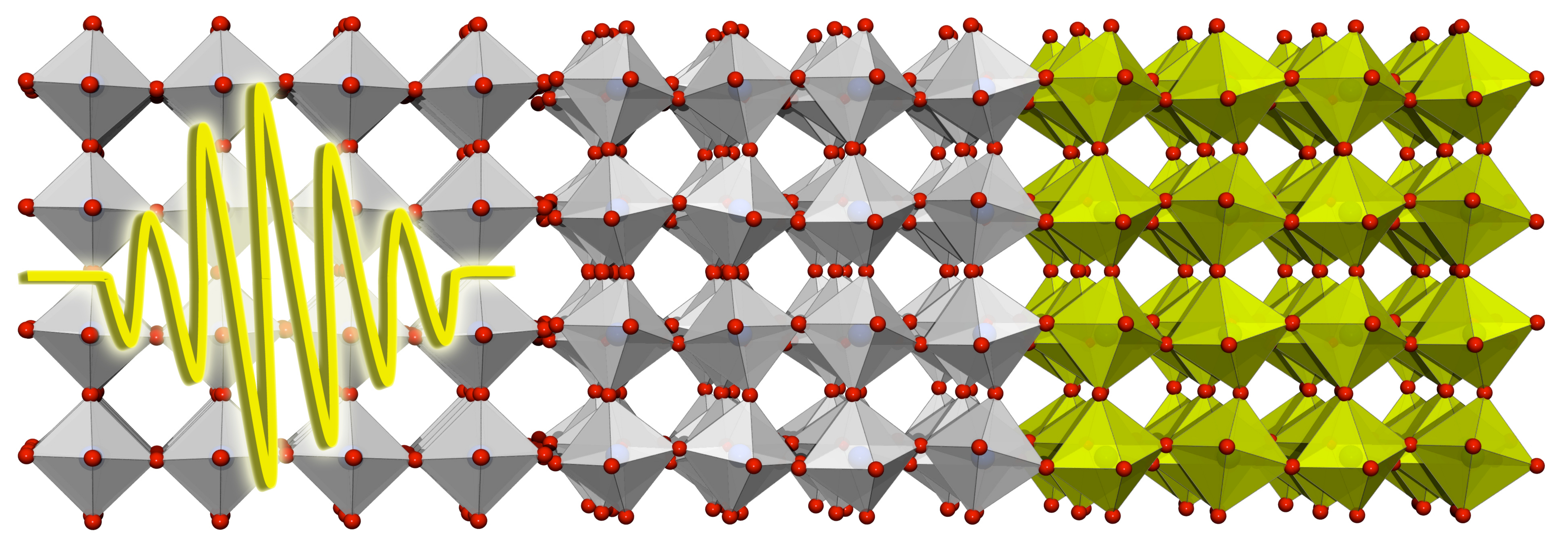}
\caption{\label{fig:subs} a, c) Right axis: relative variations in reflectivity observed 5\,ps after vibrational excitation as a function of temperature $T$ in NdNiO$_{3}$ thin films 100\,u.c. thick grown on LaAlO$_{3}$ and NdGaO$_{3}$ substrates. Solid line, left axis: dc resistivity of the same samples as a function of temperature. b, d) Photosusceptibility $\chi$ as a function of pump wavelength measured at $T=20$\,K using LaAlO$_{3}$ and NdGaO$_{3}$ substrates. The solid lines show the linear absorption  due to the infrared-active phonon of the LaAlO$_{3}$ (blue) and NdGaO$_{3}$ (red) crystals \cite{Zhang:94,0953-8984-23-4-045901}. The dashed line is the linear absorption of bulk NdNiO$_{3}$ \cite{PhysRevB.51.4830}. e) Schematic representation of the dynamic control the electronic properties of a thin film via vibrational excitation of the substrate.}
\end{center}
\end{figure}

In a first set of experiments, NdNiO$_{3}$ epitaxial thin films (100 u.c., 33\,nm thick) were deposited on (001) LaAlO$_{3}$ single crystals by off-axis RF magnetron sputtering. Note that NdNiO$_{3}$ single crystals cannot be grown in large enough size for most experiments. This substrate provides -0.5\% compressive strain to the material and reduces the metal insulator transition temperature $T_{\text{MI}}$ from a bulk value of 200~K to about 130~K, as determined through 4-point dc transport measurements (Fig. \ref{fig:subs}b). In our mid-infrared pump -  optical probe experiments the sample was excited by 150\,fs pulses tuned to photon energies between 70\,meV and 130\,meV. These pulses were generated by optical parametric amplification and difference frequency generation of the output of a Ti:sapphire laser amplifier, as detailed in Ref. \onlinecite{Manzoni:10}. Transient changes in reflectivity probed in the near (800\,nm, 1.55\,eV) and the far-infrared (1-6\,THz, 5-25\,meV) were used to characterise the electronic properties of NdNiO$_{3}$ after vibrational excitation. Broadband THz pulses were generated in a gas plasma \cite{plasma} and detected by electro-optic sampling in ZnTe and GaP crystal with 800\,nm pulses \cite{THZ} after reflection from the sample.

\begin{figure}
\begin{center}
\includegraphics[scale=0.2]{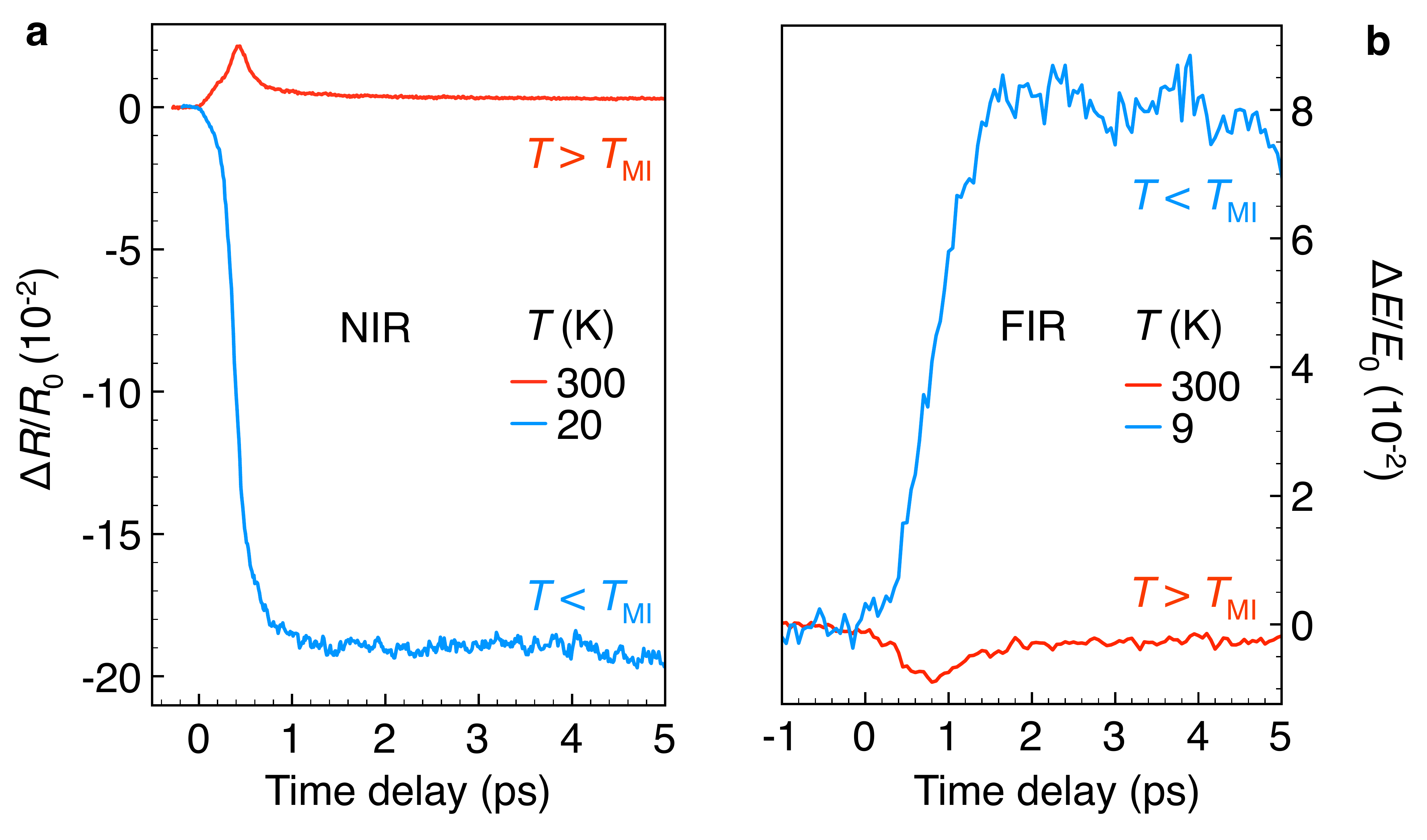}
\includegraphics[scale=0.2]{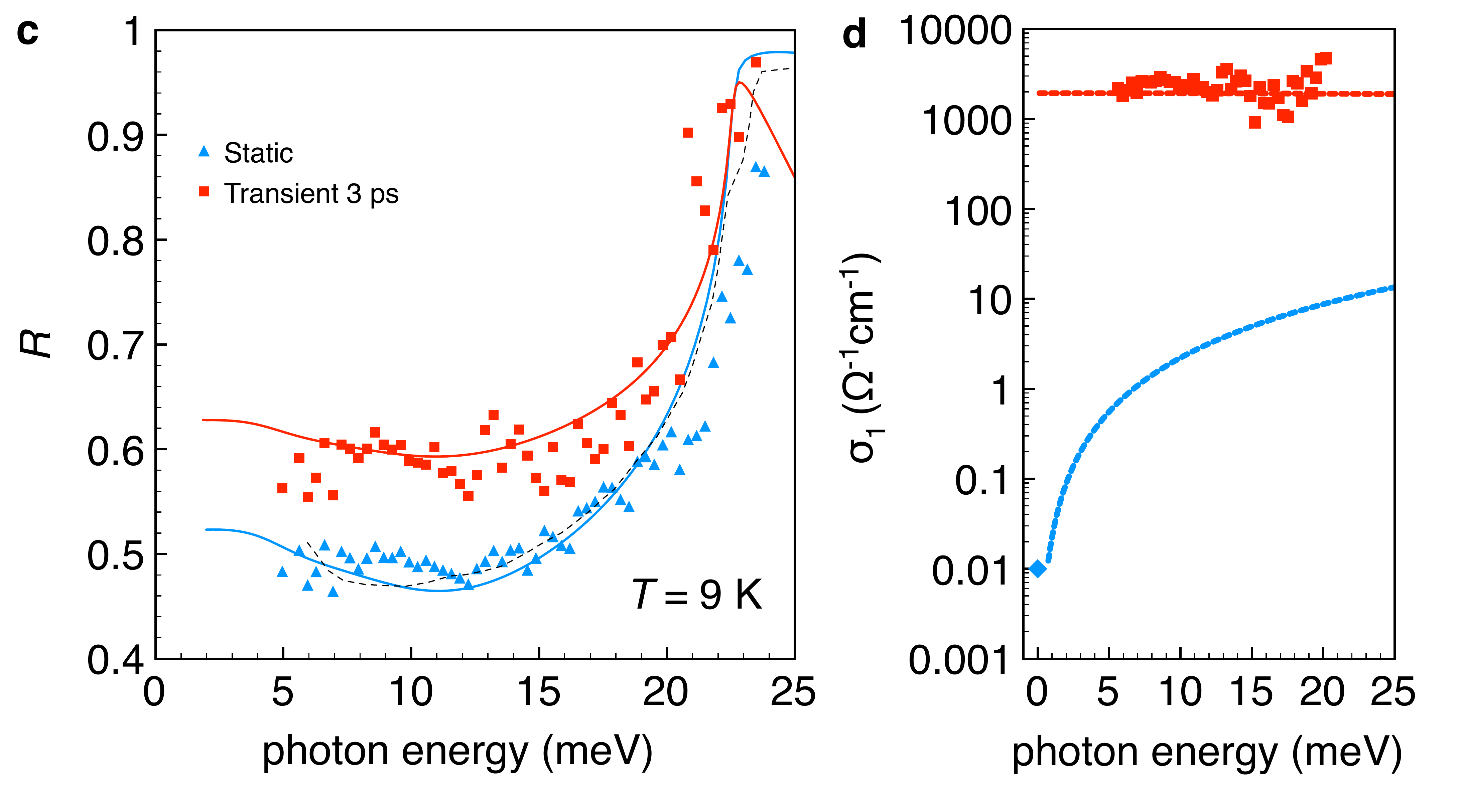}
\caption{\label{fig:deltar} a) Transient reflectivity changes at 800\,nm observed after vibrational excitation in NdNiO$_{3}$ thin films 100\,u.c. thick grown on LaAlO$_{3}$. The measurements are performed at two temperatures $T$, below and above $T_{\text{MI}}$. b) Changes of the reflected THz peak electric field (in percent) exhibiting a long-lived excited state below $T_{\text{MI}}$. c) Static (triangles) and transient (squares) THz reflectivity spectra after vibrational excitation of a NdNiO$_{3}$/LaAlO$_{3}$ heterostructure measured at $T=9$\,K. Dashed line: reflectivity of a LaAlO$_{3}$ single crystal \cite{Zhang:94}. Solid lines: calculated reflectivity of the heterostructure with NdNiO$_{3}$ in an insulating (blue) and metallic (red) state. d) Corresponding static (diamond) and transient (squares) conductivity spectra extracted for the NdNiO$_{3}$ layer from the heterostructure measured at $T=9$\,K. The static value is measured by dc transport while the transient values are extracted from the reflectivity spectra presented on the left panel. A 5 orders of magnitude modulation in dc conductivity is observed. Dashed lines: optical conductivity in the equilibrium and excited state.}
\end{center}
\end{figure}
Figure \ref{fig:deltar}a shows transient near-infrared reflectivity changes $\Delta R/R_{0}$ induced by excitation at 15\,$\mu$m wavelength for different base temperatures $T$. At room temperature, where NdNiO$_{3}$ is metallic, only a modest reflectivity increase was observed, decaying  within 2\,ps and likely associated to electronic excitations and relaxation near the Fermi level. For $T<T_{\text{MI}}$ a dramatically different response is observed. Changes in reflectivity as large as $-$20\%, persisting for hundreds of picoseconds, indicate the formation of a metastable electronic phase. This response is independent of the polarisation of either the pump or the probe beams. As evident from the temperature dependence of the photo-response (Figure \ref{fig:subs}a), such large reflectivity changes can only be achieved below the static metal-insulator-transition temperature. 

The time-dependent THz response, displayed in Figure \ref{fig:deltar}c for the spectral range between 5 and 25\,meV, reveals that this transient phase is metallic. Firstly, the frequency integrated THz reflectivity displays a long lived increase, suggestive of a phase with higher conductivity  \cite{PhysRevB.51.4830} (Figure \ref{fig:deltar}b). These changes mirror those observed when probing at 800\,nm (see above). 
Secondly, the frequency dependent THz reflectivity for the unperturbed heterostructure and for the photo-induced state are well traced by the calculated reflectivity of a 33\,nm NdNiO$_{3}$/ LaAlO$_{3}$ heterostructure with NdNiO$_{3}$ in its insulating and metallic state, respectively (Fig. \ref{fig:deltar}c). Within 3\,ps after vibrational excitation, we extract a five orders of magnitude increase in dc conductivity for the NdNiO$_{3}$. Note that for the fluence of the order 1\,mJ/cm$^{2}$ used here, and for the 4\% absorption at this wavelength, the temperature increase is estimated to be below 5\,K, pointing to a non thermal mechanism for the observed effect.

As the light induced reflectivity changes scale linearly with the pump fluence (shown below), we can define an effective photosusceptibility as $\chi=|d \rho/df|$, where $f=F(1-R)$ is the absorbed pump fluence ($F$) and $\rho=\Delta R/R_{0}$. The dependence of $\chi$ on the pump photon energy measured at $T=20$\,K is shown in Fig. \ref{fig:subs}, and highlights the most striking observation reported here. We find that the susceptibility for the photo-induced insulator-metal transition follows the  absorption of the substrate rather than that of  NdNiO$_{3}$. This phenomenon could be related to two distinct effects. One scenario involves the excitation of a substrate phonon, which may lead to a distortion coupled into the film across the interface. Alternatively, epitaxial strain may shift the resonance frequency of the thin film itself, accidentally matching the frequency of the substrate. In the first case one expects a strong dependence of the photosusceptibility on the oscillator strength of the substrate phonon. Differently, in the second case, the substrate resonance strength is not expected to influence the amplitude of the response.

To clarify this point, 100\,u.c thick NdNiO$_{3}$ epitaxial thin films deposited  on (110)\,NdGaO$_{3}$ single crystals were analysed. This substrate provides different mechanical boundary conditions (tensile strain, 1.1\%), a different vibrational resonance (74\,meV) and a much smaller oscillator strength of the phonon. NdNiO$_{3}$ thin films deposited on NdGaO$_{3}$ can be considered our best representation of the bulk material, since their transition temperature ($\sim$180\,K) is close to the one observed in ceramics ($\sim$200\,K \cite{0953-8984-9-8-003}). Figure \ref{fig:subs}c shows the temperature dependence of the 800\,nm reflectivity changes $\Delta R_{\text{5\,ps}}/R_{0}$ probed 5\,ps after vibrational excitation in NdNiO$_{3}$/NdGaO$_{3}$ heterostructures. As observed for LaAlO$_{3}$ substrates, the transient near-infrared reflectivity changes are negative, long-lived and the temperature at which the reflectivity decreases after excitation corresponds to $T_{\text{MI}}$. More importantly, the signal amplitude $\Delta R_{\text{5\,ps}}/R_{0}$ is in this case much smaller than the one observed in NdNiO$_{3}$/LaAlO$_{3}$ heterostructures. This is clearly highlighted by the dependence of the photosusceptibility on the pump photon energy shown in Figure \ref{fig:subs}d. Also for this heterostructure the photosusceptibility peaks at the phonon resonances of the substrate. Moreover, the data show a clear dependence of the photosusceptibility on the absorption coefficient of the substrate, which is three times larger for NdNiO$_{3}$/LaAlO$_{3}$. This indicates that the modulation of the electronic properties of the thin film occurs via excitation of the crystal lattice of the substrate (Figure \ref{fig:subs}e).

\begin{figure}
\begin{center}
\includegraphics[scale=0.20]{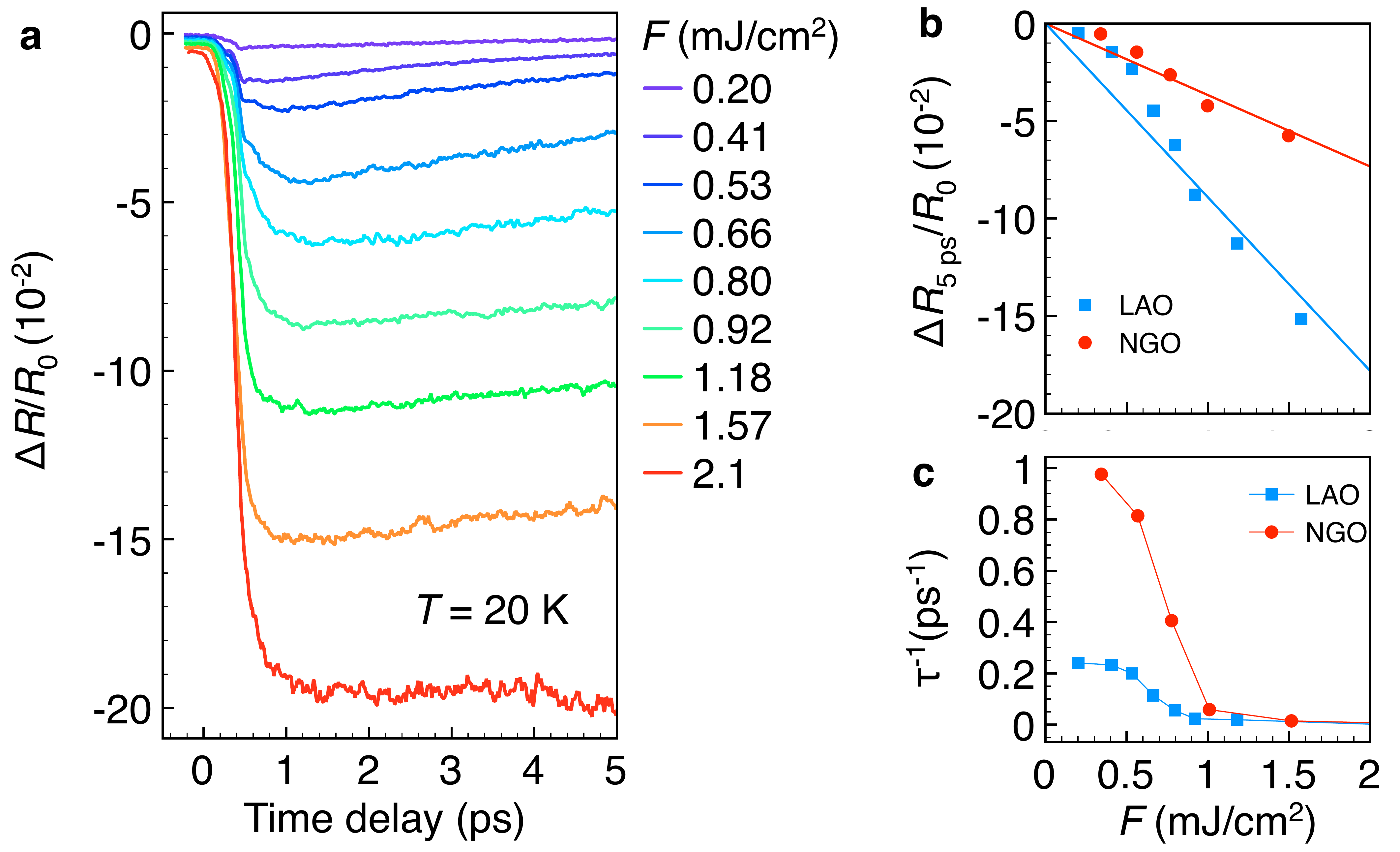}
\caption{\label{fig:flu} a) Transient reflectivity change observed after vibrational excitation in NdNiO$_{3}$ thin films 100\,u.c. thick grown on LaAlO$_{3}$. The measurements are performed at different fluences $F$. b) maximum reflectivity change $\Delta R_{\text{max}}/R_{0}$ probed at different fluences for NdNiO$_{3}$ thin films 100\,u.c. thick grown on LaAlO$_{3}$ and NdGaO$_{3}$. The solid lines are linear fit to the data used for the calculation of the photosusceptibility $\chi$.}
\end{center}
\end{figure}

Insight into the dynamics of the insulator-metal transition can be gained from the pump intensity dependence. Transient reflectivity changes measured at fluences between 0.2 and 2.1~mJ/cm$^{2}$  and $T=20$\,K, are reported in Figure \ref{fig:flu}a. The  reflectivity change $\Delta R_{\text{5\,ps}}/R_{0}$ increases linearly with the pump fluence (Figure \ref{fig:flu}b) and no saturation of the signal is observed in the experimentally accessible fluence range. As the fluence is reduced below $\sim$0.8\,mJ/cm$^{2}$ the photoinduced state decays exponentially back to equilibrium on the picosecond time scale. In this regime the relaxation time is strongly reduced with lower pump fluences (Figure  \ref{fig:flu}c). This observation can be interpreted by considering that the photo-induced transition is initiated locally, injecting a number of domains with sizes (or densities) that depend on the pump fluence. For large pump fluences the nucleated domains are large (or dense) enough to reach the percolation threshold, leading to a long lived phase. Below threshold the metallic domains never coalesce and collapse back to the parent phase. Alternatively, since the vibration of the film is driven predominantly by the excitation of the substrate, we can imagine that the amplitude of these distortions  will be maximum at the film-substrate interface, decaying into the film. Thus for low pump fluences, only the near-interface regions will be excited, whereas for higher fluences the excitation will be transferred deep into the film, stabilising the metallic state throughout the sample. Similar (but static) inhomogeneous distortion profiles induced by mechanical coupling with the substrate have recently been proposed for other perovskite heterostructures \cite{PhysRevLett.105.227203}.

In summary, this work demonstrates how optical excitation of a substrate lattice can control the electronic properties of a thin film through propagation of a dynamic distortion across the interface. This greatly broadens the scope of research of quantum phase control via optical excitation and the field of strain engineering in complex solids. The precise interaction within NdNiO$_3$ responsible for the photo-induced metallic phase remains poorly identified. One possibility is that the electronic transition is caused directly by the excitation of the Ni-O stretching mode that couples to the $B$-O bond vibration of the substrate, as changes in Ni-O bond lengths are known to occur during the thermally induced metal-insulator transition. The other possibility is that the $B$-O stretching mode couples to a rotational mode of the oxygen octahedra that would lead to changes in overlap between the Ni $d$ and O $2p$ orbitals. Such a coupling between the stretching and rotational modes has recently been observed in bulk manganites \cite{Forst:2011fk}. The most obvious way forward for this area of research is to seek conditions under which bi-directional phase control can be achieved, for example by incorporating phonon-active layers in superlattices. Understanding and engineering such couplings between structural modes in complex oxide heterostructures opens new possibilities for design and control of their scientifically fascinating and technologically useful properties.

We thank D. Nicoletti for useful discussions and J. Harms for technical support. This work was funded by the Max Planck Society and the University of Hamburg as core support for the Max Planck Research Group for Structural Dynamics. We acknowledge financial support by the SNSF through the Prospective Researcher Fellowship, the NCCR ``Materials with Novel Electronic Properties" MaNEP and Division II.

\end{document}